\documentclass[aps,amsfonts,prl,twocolumn,superscriptaddress,showpacs]{revtex4-1}
\usepackage{epsfig,amsmath,amssymb,bm,epsf,graphics,psfrag,verbatim,subfigure,color}

\begin{document}
 \title{Fiber networks below the isostatic point: fracture without stress concentration}
 \author{Leyou Zhang}
 \affiliation{Department of Physics, University of Michigan, Ann Arbor, MI 48109-1040.}
 \author{D. Zeb Rocklin}
\affiliation{Department of Physics, University of Michigan, Ann Arbor, MI 48109-1040.}
 \affiliation{Laboratory of Atomic and Solid State Physics, Cornell Univ., Ithaca, NY 14853-2501.}
 \affiliation{School of Physics, Georgia Institute of Technology, Atlanta, Georgia 30332.}
 \author{Leonard M. Sander}
 \affiliation{Physics and Complex Systems, University of Michigan, Ann Arbor, MI 48109-1040.}
 \author{Xiaoming Mao}
 \affiliation{Department of Physics, University of Michigan, Ann Arbor, MI 48109-1040.}

\date{\today}

\begin{abstract}
Crack nucleation is a ubiquitous phenomena during materials failure, because stress focuses on crack tips.  
It is known that exceptions to this general rule arise in the limit of strong disorder or vanishing mechanical stability, where stress distributes over a divergent length scale and the material displays diffusive damage.  
Here we show, using simulations, that a class of diluted lattices displays a new critical phase when they are below isostaticity, where stress never concentrates, damage always occurs over a divergent length scale, and catastrophic failure is avoided.
\end{abstract}

%
 \maketitle

When brittle materials break, they form long straight cracks; fracture occurs along a roughly planar fracture surface for three-dimensional materials or a roughly linear crack in two dimensions. This was first explained by A. Griffith ~\cite{Griffith1921}. He pointed out that  stress is \emph{concentrated} at the  tip of cracks in the material~\cite{Irwin1957}. Fracture nucleates at one of  these Griffith cracks. 
Bond failure occurs just in front of a tip, and the crack grows along a line  as a large avalanche of broken bonds.  In actual materials, disorder is always present. For large disorder, failure is spread out over a region, the process zone, whose  size we will call $\xi$ \cite{Alava2006,Bonamy2011}. In this case, successive avalanches can be small (i.e. not comparable to the system length, $L$). For $L \gg \xi$ the crack is still basically linear for two-dimensional systems.
Recently there has been much interest in models with tunable disorder for which $\xi$ can become large, and even diverge at a critical point so that material failure is delocalized~\cite{Roux1988,Hansen2003,Alava2008,Shekhawat2013,Driscoll2016}. 

In this paper, we show that in  a two-dimensional disordered fiber network model there is  a large region of parameters  where stress cannot concentrate, and damage is delocalized ($\xi\to\infty$). For the entire critical phase, there is no remnant of a Griffith crack, and avalanches are always small.

The model we consider was devised to represent disordered networks such as  biopolymer gels and some artificial porous structures. They consist of long slender fibers which are  easier to bend than to stretch~\cite{Head2003a,Broedersz2011,Broedersz2014}. The fibers are cross-linked so that we have a network with  bonds (fibers) and nodes (crosslinks).  Networks of this type  have unusual  mechanical properties such as strain-stiffening \cite{MacKintosh1995,Storm2005,Onck2005,Gardel2004,Vader2009,Broedersz2014,Feng2015,Feng2016,Licup2015,Zagar2015,Sharma2016,Feng2016} because there can be a crossover from bending-dominated  deformations at small strain  to stretching-dominated deformations at large strain. Since the fibers are easy to bend and hard to stretch the elastic modulus increases with strain. Real biopolymers such as collagen-I also show strain-stiffening and it is believed that the physical reason is the same \cite{Onck2005,Broedersz2014}.

The crossover is controlled by the \emph{central-force isostatic point} (CFIP), at which the degrees of freedom and central-force constraints balance and the system is at the verge of mechanical instability.  The CFIP occurs when $\langle z \rangle =2d$ where $\langle z \rangle$ is the average coordination number at the crosslinks and $d$ is the spatial dimension~\cite{Maxwell1864,Jacobs1995,Wyart2005a,Mao2010,Mao2011a,Ellenbroek2011,Mao2013b,Mao2013c,Lubensky2015,Zhang2015a}.  
In a network where at most two fibers meet at a crosslink, $\langle z \rangle <2d$ (dangling ends are removed because they don't contribute to elasticity) and the linear elastic moduli depend on the  bending stiffness.  Beyond the small-strain regime, force chains bearing tension emerge, the number of constraints increases, the network enters the stretching regime, and it strain stiffens.  In contrast, fiber networks with $\langle z \rangle >2d$ are stretching dominated  and do not strain stiffen (outside of the crossover region).

This paper concerns  the failure of fiber networks as  external strain is applied~\cite{Duxbury1990,Wei2016,Ovaska2017}.  For networks similar to ours  in the linear elasticity  $\xi$ diverges as the system approaches the CFIP from above; i.e. the $\langle z \rangle >2d$ side~\cite{Driscoll2016}. For this case $\xi \sim (z-2d)^{-\nu}$.  A network smaller than $\xi$ displays ``diffuse failure'' and the system breaks  when damaged regions percolate. For larger networks the crack is an effectively one- dimensional curve \cite{Shekhawat2013}.  The scale $\xi$  showed up earlier in studies of the random fuse model (RFM) in which disorder in the breaking threshold instead of proximity to the CFIP was the control parameter and the divergence of $\xi$ occurred for infinite disorder~\cite{Hansen1991,Alava2008,Shekhawat2013}.  
For both types of model the system generically flows to the nucleation fixed point when $L \gg \xi$.

In this paper we address the breaking of a model fiber network \emph{below} the CFIP ($\langle z \rangle <2d$). The system is bending dominated for small strain. To break bonds we must go into the non-linear regime (strain-stiffening), so the entire process is controlled by the forming force chains. The system shows remarkable behavior: in the finite parameter range  $\langle z \rangle <2d$ (but substantially above geometric percolation), the process zone diverges, $\xi\to\infty$, and this fiber network shows diffuse failure even in the thermodynamic limit.  
The system is so disordered that the classic Griffith scenario of breaking near a crack tip is never relevant. Stress concentration is overwhelmed by disorder.

\begin{figure}[h]
  \includegraphics[width=0.5\textwidth]{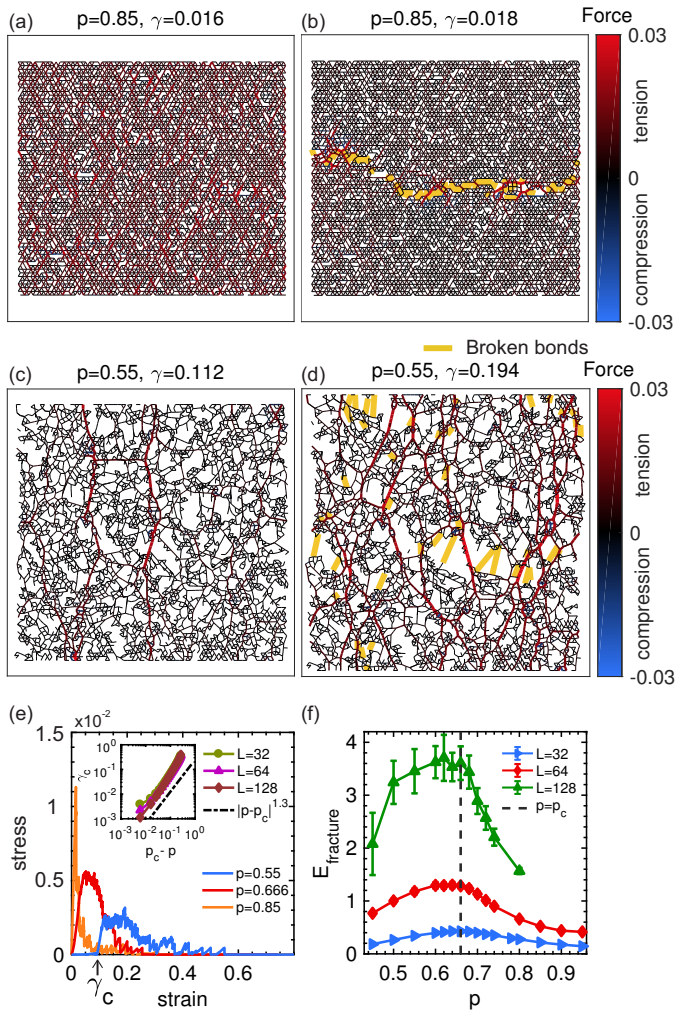}
  \caption{(Color online)  (a) Lattice with $p=0.85, \gamma=0.016$: stress concentration near crack tips.  (b) $p=0.85, \gamma=0.02$: a Griffith crack.  (c) $p=0.55, \gamma=0.112$: force chains emerge. (d)  $p=0.55, \gamma=0.194$: diffuse failure. 
  Bonds are colored according to force; broken bonds are also marked (see legend). Bonds with large force are thicker.
(e) Stress-strain curves at various $p$ for $L=128$. The critical strain, $\gamma_c$, is labeled for $p=0.55$. Inset: Scaling relation $\gamma_c \sim (p_c-p)^{\beta}$ with $\beta\simeq 1.3$~\cite{Feng2016}, for different $L$. (f) The fracture energy reaches a maximum when $p\simeq p_c$. }
\label{fig:lattice}
\end{figure}

Our model is an $L\times L$ diluted triangular lattice of springs which all have the same spring constant and each bond is present with probability $p$ \cite{Das2007,Broedersz2011,Mao2013b,Mao2013c,Feng2015,Feng2016}.  The elastic energy of the system is the sum of the stretching energies of all the bonds.  We neglect bending energy entirely since we are interested in the nonlinear regime.  In this case, the network exhibit a   continuous rigidity percolation transition at $p_c\simeq0.6602$ (very close to the mean-field value of $p=2d/6=2/3$ which is the CFIP)~\cite{Jacobs1995}.  
The effect of adding bending energy will be discussed in the end.  In the initial state all bonds are at their rest length $a$.

 We apply a uniaxial strain $u_{yy}=\gamma$ and minimize the system's elastic energy using the FIRE algorithm \cite{Bitzek2006}. Periodic boundary conditions are applied for the $x$-direction, while the top and bottom boundaries are held as rigid bars to impose strain. 
When a bond is stretched beyond $(1+\lambda)a$, it is broken, i.e., removed.  Here $\lambda$ is a threshold parameter which is  taken to be $0.03$ for all the bonds. We re-equilibrate the node positions after each bond breaks.  The breaking of a bond can  trigger other breaking events (an avalanche). We continue the process until no more bonds are beyond threshold. Then we  increase the strain by a  step small enough that at most one bond is beyond threshold  (though the subsequent avalanche  can involve multiple bonds breaking at the same time).  The strain is increased until the lattice is broken into two disconnected pieces (final failure).
To make contact with the Griffith crack regime, a small notch of 8 broken bonds is placed in the center of the lattice  to nucleate a crack. 
We scan the  size of the system from $L=32$ to $256$ and $p$ from 0.5 to 0.85. 

Examples are shown in Fig.~\ref{fig:lattice}(a-d). The $p>p_c$ network is well described by the Griffith theory.  
When $p<p_c$,  the system is bending dominated at small strain.  We have zero bending stiffness; thus  the network develops no stress until a critical strain $\gamma_c \sim (p_c-p)^{\beta}$ with $\beta\simeq 1.3$ \cite{Feng2016}.  Beyond $\gamma_c$ bonds start to be stretched and will break when they extend beyond the threshold length; see Fig.~\ref{fig:lattice}(c-e).  The total energy absorbed during the fracturing process, the fracture energy, as shown in  Fig.~\ref{fig:lattice}(f), peaks at $p\simeq p_c$.  This indicate that networks close to the CFIP display highest toughness.  

\begin{figure}
  \includegraphics[width=0.5\textwidth]{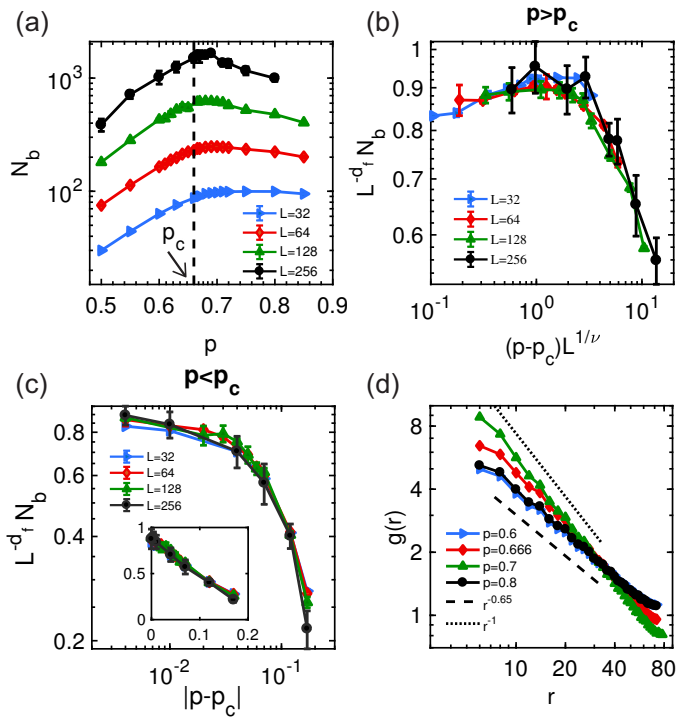}
  \caption{(Color online) (a) $N_b$ as a function of $p$ for $L=32, 64, 128, 256$, displaying a maximum at $p_c$. (b) The pair correlation function $g(r)$ of broken bonds at $L=128$.  (c) Scaling collapses of $N_b$ for $p<p_c$ with $\nu\to\infty$. (inset) The same plot in a linear-linear scale. (d) Scaling collapses of $N_b$ for $p>p_c$ with $\nu=1.21$.  
  }
\label{fig:Nb}
\end{figure}

To extract the length scale, $\xi$, we propose a standard finite-size scaling form for the total number of broken bonds, $N_b$, at final failure (Fig.~\ref{fig:Nb}).  Two simple limits of $N_b$ are that $N_b\sim L$ as $p\to 1$ (crack nucleation) and $N_b \to 0$ as $p \to p_{\textrm{GP}}\simeq 0.347$ (geometric percolation~\cite{Sykes1964}).  In the intermediate range close to the CFIP and our data can be collapsed using the following form:
\begin{align}
	N_b = L^{d_f} \mathcal{N}\left(  \vert p-p_c \vert L^{1/\nu}\right) ,
\end{align}
with $d_f=1.35$.  
We find a good collapse for $p>p_c$ using $\nu=1.21$, the value from rigidity percolation~\cite{Jacobs1995,Jacobs1996}: see Fig.~\ref{fig:Nb}(b).  This is consistent  with a correlation length $\xi \sim  \vert p-p_c \vert^{-\nu}$.  From earlier studies we know the physical meaning of $\xi$: It is the scale at which the probability for a network being rigid exhibits significant fluctuations.  Here it controls the size  of the  process zone, because \emph{below this scale stress cannot be concentrated at the crack tip}.

For $p<p_c$ we find that the data collapses using $\nu\to\infty$, Fig.~\ref{fig:Nb}(c), indicating an infinite correlation length -- a breakdown of finite size scaling.  Networks below the CFIP break when the strain exceeds $\gamma_c$ (onset of strain stiffening), deep in the regime of nonlinear elasticity.  In these networks a dynamic steady state occurs during fracture: force chains emerge, then break,  and new force chains emerge.  The system constantly drives itself to the CFIP until  final failure.  \emph{The length scale of the process zone is always $\xi\to\infty$ for $p<p_c$}. We never have a localized crack  even as $L\to\infty$.  Furthermore, the collapsed curve suggests that for $p<p_c$ the scaling function $\mathcal{N}$ takes the form $\mathcal{N}(y)=a_1-a_2y$ where $a_1\simeq0.9$ and $a_2\simeq3.8$ are constants. 
The different scalings above and below the CFIP suggests that for large $L$, the decrease of $N_b$ as $p$ deviates from $p_c$ is much slower on the $p<p_c$ side than the $p>p_c$ side.  In other words, networks at $p<p_c$ are highly dissipative during the fracture. This is prominent in the fracture energy plot [Fig.~\ref{fig:lattice}(f)].

We interpret the exponent $d_f=1.35$ as the fractal dimension of the cluster of broken bonds in the  process zone.  To verify the geometric interpretation of $d_f$, we measured the pair correlation function $g(r)$ [Fig.~\ref{fig:Nb}(d)].  For $p<p_c$ the process zone is the whole lattice and  we observe $g(r)\sim r^{d_f-2}$.  This crosses over to $g(r)\sim r^{-1}$ for $p \to 1$ as a result of crack nucleation.  To get $g(r)$ we used the positions of the broken bonds in the undeformed state instead of their position at breaking because the deformed state is constantly evolving during strain and does not provide a well-defined metric.  We disregard the first 20\% of broken bonds for $p<p_c$ to eliminate uncorrelated damage at the beginning \cite{Nukala2005,Nukala2004}. 

\begin{figure}
    \centering
    \includegraphics[width=0.5\textwidth]{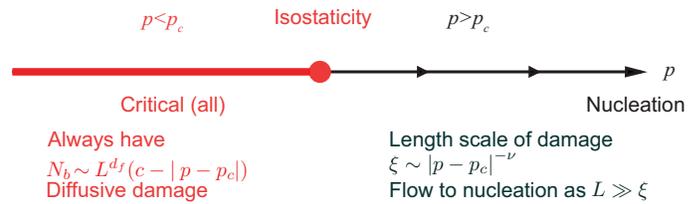}
    \caption{(Color online) Schematic phase diagram. Arrows give the direction of flow as $L\to\infty$.}
    \label{fig:diagram}
\end{figure}

Our scaling collapse of $N_b$ leads to the phase diagram shown in Fig.~\ref{fig:diagram}, 
with the arrows representing parameter flow on coarse-graining.  For $p>p_c$ the system always flows to the nucleation fixed point as $L/\xi$ increases, whereas for $p<p_c$ the system does not flow ($\xi\to\infty$). Rather it drives itself to the CFIP in the nonlinear regime during  fracture.  We call this regime the ``critical phase''.  This phase shares interesting similarities with the phenomena that molecular motor activities drive biopolymer gels to a critically connected state~\cite{Alvarado2013}.

\begin{figure}
  \includegraphics[width=0.5\textwidth]{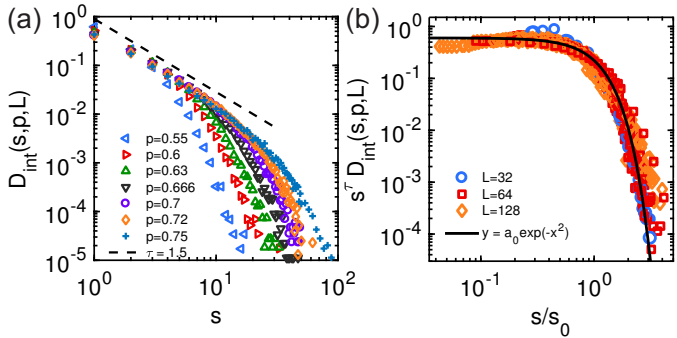}
  \caption{(Color online) (a) Integrated avalanche size distribution $D_{int}(s,p,L)$,  $L=128$, various $p$. Dashed line: the power law $s^{-\tau}$,  $\tau=1.5$.
(b) The collapse of $D_{int}(s,p,L)$ for systems of $p$ from 0.55 to 0.75 and $L = 32,64,128$, following Eq.\eqref{EQ:s0}.   Solid line: empirical function that fits the master curve.
}
  \label{fig:aval}
\end{figure}

To characterize avalanches in these fiber networks and verify the scaling scenario we proposed based on $N_b$, we also studied the integrated  size distribution $D_{int}(s, p, L)$  for all avalanches of size $s$ until failure.  
A standard form from the literature \cite{Sethna2001,Nukala2005,Alava2006} is  $D_{int}(s, p, L) = s^{-\tau} \mathcal{D}(s/s_0)$ where $s_0$ is a cutoff size for the power law $s^{-\tau}$. 
As shown in Fig.~\ref{fig:aval}(a) for our model $\tau=3/2$.

The cutoff size $s_0$ is a function of $p$ and $L$, and the scaling we obtained above, $\xi\sim \vert p-p_c \vert^{-\nu} $,  provides a way to collapse $D_{int}(s, p, L)$ onto a master curve (Fig.~\ref{fig:aval}b) using the following form
\begin{align}\label{EQ:s0}
	s_0 (p,L) = L^{D} \mathcal{S} (\vert p-p_c \vert L^{1/\nu}).
\end{align}
For our model the data is consistent with $D \approx 1$. This should be compared to random fuse models \cite{Arcangelis1989,Nukala2005} for which $D \approx 1.1$. Also,
$\mathcal{S}(x)=c_0 + c_1 x$ for $p>p_c$ and $\mathcal{S}(x)=c_0 + c_2 x$ for $p<p_c$, where $c_0\simeq0.11$, $c_1\simeq0.024$, and $c_2 \simeq -6.1$. 
We have used the correlation length exponent we obtained from the scaling of $N_b$, namely $\nu=1.21$ for $p>p_c$ and $\nu\to\infty$ for $p<p_c$.  This good collapse is consistent with our interpretation of $\xi$ and thus the phase diagram in Fig.~\ref{fig:diagram}.

\begin{figure}
  \includegraphics[width=0.5\textwidth]{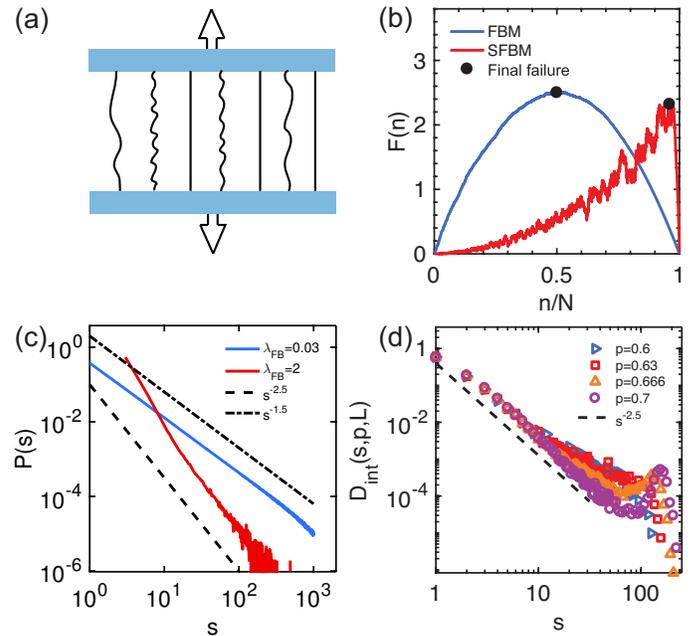}
  \caption{(a) SFBM schematic.  (b) Force as a function of number of broken fibers in  FBM and  SFBM.  
  (c) Crossover of $\tau$ in the SFBM with  $\lambda$.  (d) $D_{int}(s,p,L), \lambda=0.2$  various $p$ for the fiber network model.
}
  \label{fig:lambda}
\end{figure}

To provide an intuitive picture for the force chain forming-breaking steady state and the $\tau=3/2$ in our fiber networks, we introduce a toy model,  the ``slack fiber bundle model'' (SFBM, see Fig.~\ref{fig:lambda}a) inspired by the fiber bundle model (FBM)  for  fracture with random breaking thresholds~\cite{Hemmer1992,Alava2006,Pradhan2010}.  In the FBM, two plates are connected by  fibers with randomly chosen breaking thresholds.  The plates are pulled apart with force $F$, which is equally shared by all fibers.  Failure in the FBM includes avalanches where the breaking of one fiber makes others break. The avalanche size distribution  is equivalent to the first return time of a biased random walk \cite{Hemmer1992,Sornette1992}:
\begin{align}
	D^{FBM}(s,F) \sim s^{-3/2} e^{-s/s_0(F)},
\end{align}
where the bias (the ratio between mean and variance) is $b=s_{0}(F)^{-1/2}\sim F_c-F$, where $F_c$ is the critical force where the final catastrophic failure occurs.  The integrated avalanche distribution over the whole process is: 
\begin{align}
	D^{FBM}_{int}(s,F) \sim \int_{0}^{F_c} D^{FBM}(s,F) dF \sim s^{-5/2}.
	\end{align}
The $5/2$ exponent is the result of vanishing bias (divergent cutoff $s_0$) as the final failure is approached (see Fig.~\ref{fig:lambda}(b) and SM), and is characteristic of most brittle fracture processes.

In the SFBM, instead of a distribution of the threshold, we assume a distribution of fibers' rest length, $P(x_0)$, and assume that the fiber will break when it is stretched beyond $(1+\lambda)x_0$.  Thus, the load $F$ is not equally shared by all fibers. Instead,  fibers with rest length longer than the distance between the plates remain slack until the distance between plates increases to their $x_0$.  In this model (as in the original 2D network) new force chains constantly emerge in the process of failure. We assume the distribution of $x_0$ is quite random, i.e. the standard deviation  is comparable to the mean. It is  shown in detail in the SM   that the bias in the random walk of force is a constant (Fig.~\ref{fig:lambda}b), $s_0^{-1/2}\sim \lambda $, if $\lambda\ll 1$.  The integrated avalanche size distribution is:
\begin{align}
	D^{SFBM}_{int}(s,F) \sim \int_{0}^{F_c} D^{SFBM}(s,F) dF \sim s^{-3/2}e^{-\lambda^2 s/2}.
\end{align}
This is valid when  $\lambda\ll 1$, meaning that only a small fraction of the fibers are stretched at any given force. 
The failure process in the SFBM is a steady state where new fibers join the load-bearing group and ones beyond threshold break. This steady-state process with $\tau = 3/2$ is reminiscent of other mean-field "self-organized branching processes" such as plastic slip events~\cite{zapperi1995self, dahmen2009micromechanical,Dahmen2011}. In contrast, in  the FBM, all of the fibers are stretched, and the fracture process evolves significantly, culminating in a catastrophic failure.

The SFBM  crosses over to  FBM behavior when $\lambda\gtrsim 1$.  In this case, a macroscopic fraction of the fibers are stretched at the same time,  the distribution of rest length is overwhelmed by the distribution of the threshold, and the avalanches are like those in the FBM.  We find that in the SFBM, taking large $\lambda$ leads to a crossover from $\tau=3/2$ to $5/2$ (Fig.~\ref{fig:lambda}c).  The same phenomena occur in the fiber networks as well, as shown in Fig.~\ref{fig:lambda}(d). With large $\lambda$ the exponent is 5/2, and we have localized crack nucleation,  destroying the critical phase for $p<p_c$.  

In summary, we investigated failure in fiber networks for which nonlinear elasticity dominates. We found that the system drives itself to criticality with $\xi\to\infty$ (diffuse failure) for  $p<p_c$.  The avalanche size distribution agrees with our interpretation. We also proposed a toy model for the avalanche sizes, the SFBM, which gives insight into the differences between the conventional (localized crack) case and our diffuse failure model.

It is natural to ask whether our scaling analysis applies to real polymer gels.  Real biopolymers exhibit bending stiffness  which we ignore.  Also, the failure of individual polymers which involves breaking down complicated macromolecular structures~\cite{Buehler2006}, is much more complicated than the simple threshold  we used.  Both these effects make it more likely for the network to focus stress and crossover to crack nucleation in the thermodynamic limit~\cite{Ovaska2017,Wei2016}.  The critical regime we found for $p<p_c$  may be observable in man-made materials in which bending stiffness and threshold are controlled to be very small.  The highly dissipative nature of the process for $p<p_c$ may allow the design of materials which absorb  a large amount of energy when they break~\cite{Sun2012a}.

\emph{Acknowledgments}  We acknowledge informative discussions with James Sethna.  This work was supported in part by the National Science Foundation Grant No. NSF DMR-1609051 (XM and LZ), the ICAM postdoctoral fellowship, the Bethe/KIC Fellowship, and the National Science Foundation Grant No. NSF DMR-1308089 (DZR).

\appendix
\section{Supplementary Materials: The slack fiber bundle model (SFBM)}
In the main text we introduced the SFBM to characterize the dynamic steady state in the fracturing of the fiber networks below the central-force isostatic point (CFIP).  In the SFBM, there are $N$ fibers between two parallel plates, and an external mechanical load cause the fibers to break in a series of avalanches.  

The SFBM is inspired by the fiber bundle model (FBM)~\cite{Hemmer1992,Alava2006} which has been used to describe brittle failure.  In the FBM,  the $N$ fibers all have the same rest length but a distribution of breaking threshold.  As we discuss in the main text, avalanches in the FBM are described by the first return time in the biased random walk of the external force as a function of the number of broken fibers, and the bias of this random walk vanishes as the catastrophic failure is approached, leading to an avalanche exponent of $5/2$.

In the FBM, instead of breaking threshold, the rest length $x_0$ of the fibers obey a probability distribution $P(x_0)$, and they break when stretched to length $(1+\lambda)x_0$. So the force on a fiber, when the distance between the two plates is $x$, is given by
\begin{equation}
    \begin{aligned}
    f(x) = 
    \begin{cases}
      k(x-x_0), & \text{if}\ x_0\le x \le (1+\lambda) x_0\\
      0, & \text{otherwise.}
    \end{cases}
    \end{aligned}
\end{equation}When there are $N$ fibers in the system, the system's load is $F(x) = \sum_{i=1}^N f_i(x)$. 
As the two plates are stretched apart, new fibers join the stretched, load-bearing group, and fibers stretched beyond their threshold break.  
Following the analysis of the FBM, we consider the random walk of the total force $F_i$ when the $i$-th fiber is about to break.  Although the distance between the plates monotonically increases, the force the fibers can support can decrease, leading to avalanches.  Thus the avalanche size distribution maps to the first return time of the biased random walk of $F_i$.  To obtain this distribution we need the mean $\mu=\langle F_{i+1}-F_i\rangle$ and variance $\sigma^2=\langle (F_{i+1}-F_i)^2\rangle-\mu^2$ of this random walk.  Because the force difference right before the $i$-th and the $i+1$-th breaking is
\begin{align}\label{EQ:dF}
	F_{i+1}-F_i =& k\Big\lbrace  -\lambda x_{0,i} + (1+\lambda) (x_{0,i+1} -x_{0,i} ) N P( x_{0,i}) \nonumber\\
	&\times\lbrack(1+\lambda ) x_{0,i} - x_{0,i+1}\rbrack \Big\rbrace ,
\end{align}
where the first term is the decrease of force due to the $i$-th fiber breaking, and the second term is the increase of force because the load bearing group are stretched $(1+\lambda) (x_{0,i+1} -x_{0,i} ) $ more.  The number of fibers in the load bearing group is calculated from the product of the probability distribution and the range of rest length of fibers belonging to this group (must be stretched at distance $(1+\lambda) x_{0,i}$ and not broken before $i+1$-th fiber break).  We ignore the fibers that start to join the load-bearing group, because the stretch on these fibers is of higher order in $(x_{0,i+1} -x_{0,i} )$.

Here we take the approximation that $\lambda\ll 1$,  $N\gg 1$ and $\lambda x N P(x) \gg 1$ so that at any time, a large number of fibers are in the load-bearing group [which also means that the rest length difference between $i$-th and $i+1$-th fibers, $\delta x_0\equiv x_{0,i+1} - x_{0,i}\sim 1/NP(x)$, is much smaller than the stretching range of each fiber $\lambda x$, and guarantees the positiveness of the last term in Eq.~\eqref{EQ:dF}].

This can be further simplified into
\begin{align}
	F_{i+1}-F_i = k \lambda x_{0,i}\left\lbrack  -1 + (1+\lambda) N P( x_{0,i}) \delta x_0 \right\rbrack ,
\end{align}
where we have dropped a term of higher order in $\delta x/(\lambda x)$.  

Next we consider the mean and variance of this quantity.  Realizing that the only quantity that is stochastic in this expression is $\delta x_0$, the ``step size'' between fibers when they are ordered according to their rest length.   (The value of $x_{0,i}$ is given for given $i$.)
For the mean, it is straightforward that 
\begin{align}
	N P( x_{0,i}) \langle\delta x_0\rangle=1,
\end{align}
 leading to
\begin{align}
	\mu=\langle F_{i+1}-F_i\rangle = k \lambda^2 x_{0,i} .
\end{align}
The variance involves  $\langle\delta x_0^2\rangle$.  Using properties of Poisson process we have 
\begin{align}
	\langle\delta x_0^2\rangle = \frac{2}{(N P( x_{0,i}) )^2},
\end{align}
and thus
\begin{align}
	\sigma^2=\langle(F_{i+1}-F_i)^2\rangle-\mu^2 = k^2 \lambda^2 x_{0,i}^2 .
\end{align}

 It is well understood that the first return time of a biased random walk follows a power law of exponent $-3/2$ multiplying an exponential cut-off $\exp^{-s/s_0}$, and the cut-off $s_0 = 2\sigma^2/\mu^2$\cite{Pradhan2010}. Therefore, the avalanche size distribution of the SFBM is
 \begin{equation}
 P(s) \sim s^{-3/2} e^{-{\lambda^2}{s}/2}.
\end{equation}

Thus we see that our SFBM displays a power-law distribution of avalanches with an exponential cutoff determined by the fibers' breaking threshold. Unlike the original FBM, which has a 5/2 exponent, the SFBM displays the 3/2 behavior characteristic of a steady-state process, since new fibers tighten and begin to bear stress to compensate for the others that break.

\end{document}